\documentclass[11pt]{article}

\newsavebox{\junk}
\savebox{\junk}[1.6mm]{\hbox{$|\!|\!|$}}

\def\bbbz{{\mathchoice {\hbox{$\sf\textstyle Z\kern-0.4em Z$}}
{\hbox{$\sf\textstyle Z\kern-0.4em Z$}}
{\hbox{$\sf\scriptstyle Z\kern-0.3em Z$}}
{\hbox{$\sf\scriptscriptstyle Z\kern-0.2em Z$}}}}
\def\sq{\hbox{\rlap{$\sqcap$}$\sqcup$}}

\usepackage{amsmath} 
\usepackage{amsfonts}
\usepackage{latexsym}
\usepackage{bm}
\usepackage{multirow}

\newcommand{\ben}{\begin{enumerate}}
\newcommand{\een}{\end{enumerate}}
\newcommand{\bit}{\begin{itemize}}
\newcommand{\eit}{\end{itemize}}

\newtheorem{theorem}{Theorem}[section]
\newtheorem{proposition}[theorem]{Proposition}

\newcommand{\ba}{\begin{array}{rcl}}
\newcommand{\ea}{\end{array}}
\newcommand{\bt}{\begin{theorem}}
\newcommand{\et}{\end{theorem}}
\newcommand{\bd}{\begin{description}}
\newcommand{\ed}{\end{description}}

\def\slabel#1{\label{s:#1}}

\def\elabel#1{\label{e:#1}}
\def\eq#1/{(\ref{e:#1})}

\def\Section#1{Section~\ref{s:#1}}

\def\beq{\begin{equation}}
\def\eeq{\end{equation}}
\def\beqa{\begin{eqnarray}}
\def\eeqa{\end{eqnarray}}

\def\qed{\ifmmode\sq\else{\unskip\nobreak\hfil
\penalty50\hskip1em\null\nobreak\hfil\sq
\parfillskip=0pt\finalhyphendemerits=0\endgraf}\fi}

\def\sqr#1#2{{\vcenter{\hrule height.#2pt
      \hbox{\vrule width.#2pt height#1pt \kern#1pt
         \vrule width.#2pt}
       \hrule height.#2pt}}}

\newcommand{\bc}{\begin{corollary}}
\newcommand{\ec}{\end{corollary}}

\newcommand{\bp}{\begin{proposition}}
\newcommand{\ep}{\end{proposition}}

\def\eye(#1){{\bf (#1)}\quad}

\def\taboo#1{{{}_{#1}}}

\def\0P{\taboo{0}P}
\def\0Pn{\taboo{0}P^n}

\usepackage{bbm}
\usepackage{amsmath}
\DeclareSymbolFont{largesymbols}{OMX}{yhex}{m}{n}
\DeclareMathAccent{\widewidehat}{\mathord}{largesymbols}{"62}

\usepackage{color}
\usepackage{graphicx}
\usepackage{subcaption}
\usepackage{caption}
\usepackage{tikz}
\usepackage[version=4]{mhchem}
\usepackage{amsmath,amssymb,tikz-cd}
\usetikzlibrary{matrix}

\usepackage[ruled,vlined,algosection,algo2e]{algorithm2e}
\newtheorem{algorithm}[theorem]{Algorithm}

\usepackage[labelformat=simple]{subcaption}

\DeclareMathAlphabet{\mathpzc}{OT1}{pzc}{m}{it}

\usepackage{algorithm}
\usepackage{algorithmic}

\date{November 12, 2021}

\title{Controlled Accuracy Gibbs Sampling of Order Constrained Non-IID Ordered Random Variates}

\def\jem{Postal Address:
      Department of Applied Mathematics,
      University of Colorado, Box 526
      Boulder CO 80309-0526, USA; email: corcoran@colorado.edu, 
      phone: 303-492-0685}

\author {J. N. Corcoran, C. Miller\\
University of Colorado \thanks{\jem} }

\begin{document}

%\pagebreak
%\tableofcontents
%\pagebreak

\maketitle
\vspace{-.8cm}

\begin{abstract} \small \noindent
Order statistics arising from $m$ independent but not identically distributed random variables are typically constructed by arranging some $X_{1}, X_{2}, \ldots, X_{m}$, with $X_{i}$ having distribution function $F_{i}(x)$,  in increasing order denoted as $X_{(1)} \leq X_{(2)} \leq \ldots \leq X_{(m)}$. In this case, $X_{(i)}$ is not necessarily associated with $F_{i}(x)$. Assuming one can simulate values from each distribution, one can generate such ``non-iid" order statistics  by simulating $X_{i}$ from $F_{i}$, for $i=1,2,\ldots, m$, and arranging them in order. In this paper, we consider the problem of simulating ordered values $X_{(1)}, X_{(2)}, \ldots, X_{(m)}$ such that the marginal distribution of $X_{(i)}$ is $F_{i}(x)$. This problem arises in Bayesian principal components analysis (BPCA) where the $X_{i}$ are ordered eigenvalues that are a posteriori independent but not identically distributed. We propose a novel coupling-from-the-past algorithm to ``perfectly" (up to computable order of accuracy) simulate such {\emph{order-constrained non-iid}} order statistics. We demonstrate the effectiveness of our approach for several examples, including the BPCA problem.

\bigskip

\end{abstract}


\setcounter{page}{0}

%\tableofcontents

%%%%%%%%%%%%%%%%%%%%%%%%%%%%%%%%%%%%%%%%%%%%%%%%%%%%%%%%%%%%%%%%%%%%%%%%%
%%%%%%%%%%%%%%%%%%%%%%%      INTRODUCTION       %%%%%%%%%%%%%%%%%%%%%%%%%
%%%%%%%%%%%%%%%%%%%%%%%%%%%%%%%%%%%%%%%%%%%%%%%%%%%%%%%%%%%%%%%%%%%%%%%%%

\section{Introduction}
\slabel{int} 
Suppose that $X_{1}, X_{2}, \ldots, X_{m}$ are independent and identically distributed (iid) random variables from a distribution with probability density function (pdf) $f$. The {\emph{order statistics}} are defined and denoted as $X_{(1)},X_{(2)}, \ldots, X_{(m)}$ where $X_{(1)} = \min (X_{1}, X_{2}, \ldots, X_{m})$, $X_{(m)} = \max (X_{1}, X_{2}, \ldots, X_{m})$, and, in general, $X_{(i)}$ is the $i$th smallest value from the sample $X_{1}, X_{2}, \ldots, X_{m}$. Assuming that one can simulate values from the distribution with pdf $f$, one can simulate order statistics by simulating $m$ iid values from $f$ and then arranging them order. In the case of {\emph{independent and non-identically distributed}} ({\emph{inid}}) random variables, where $X_{i}$ comes from a distribution with pdf $f_{i}$, order statistics $X_{(1)}, X_{(2)}, \ldots, X_{(m)}$ are still defined in the same way. One can simulate these {\emph{inid order statistics}} by independently simulating $X_{i}$ from the distribution with pdf $f_{i}$ and then arranging them in order. In this case, the $i$th order statistic is not necessarily the value simulated from the $i$th distribution. Imposing such a constraint makes methods of simulation much more challenging and is the subject of this paper. We will call such random variates {\emph {order constrained  non-identically distributed}} ({\emph{OCNID}}) order statistics.

The problem of simulating ordered values corresponding to specific non-identical distributions often comes up in the context of Bayesian principal components analysis (BPCA) \cite{kazianka2016} \cite{minka2001} \cite{suarez2017} \cite{zhang2004b}, where it is of interest to be able to simulate ordered eigenvalues from a posterior distribution. We will discuss this particular problem in more detail and give an example in \Section{bpca}. 

In this paper, we will use the idea of {\emph{perfect simulation}} or {\emph{coupling-from-the-past}} (CFTP) in order to simulate $X_{1}, X_{2}, \ldots, X_{m}$ from a joint density $f(x_{1}, x_{2}, \ldots, x_{m})$ of the form
\beq
\elabel{target}
f(x_{1}, x_{2}, \ldots, x_{m}) \propto \left[ \prod_{i=1}^{m} f_{i}(x_{i}) \right] \, I_{\{x_{1}<x_{2}<\cdots < x_{m}\}}
\eeq
where $I_{\{x_{1}<x_{2}<\cdots < x_{m}\}}$ is an indicator taking the value $1$ if $x_{1}< x_{2}< \cdots < x_{m}$ and zero otherwise. The $f_{i}$ are pdfs that, in BPCA, are typically from the same family of distributions, often differing through a scale parameter. However, for the algorithm we present in this paper they need not be related in any way other than by having the same support. We will refer to the joint density given by (\ref{e:target}) as the ``target density".

We now briefly review the concept of perfect simulation and describe a novel ``perfect" Gibbs sampler to sample exactly from (\ref{e:target}) under the assumption that one can simulate values from each $f_{i}$ using the {\emph{inverse-cdf method}} which requires that one can write down and invert the cumulative distribution function (cdf) associated with $f_{i}$. In \Section{bpca} we apply our algorithm to distributions where the inverse cdf is not available but is estimated numerically. We wish to point out that our  algorithm is actually ``$\varepsilon$-perfect'' in the sense that, while we will not get exact draws (sampled values) from (\ref{e:target}), we can get arbitrarily close as described in \Section{gibbs}. Indeed the draws can be perfect with respect to any given machine precision.

\section{The Perfect Gibbs Sampler}
\slabel{gibbs}
 
A Markov chain Monte Carlo (MCMC) algorithm allows us to create a Markov chain that converges to a draw from the target distribution given by (\ref{e:target}). Perfect simulation (sampling) algorithms comprise a subclass of MCMC algorithms that are free of convergence error. The idea of perfect simulation is to find a random time $-T$, known as a {\emph{backward coupling time}} (BCT), in the past such 
that, if we construct sample paths for the chain  
starting at time $-T$ for every possible starting point,  all paths will have come together or ``coupled" by time zero. The common value of the paths at time zero is a
draw from the limiting distribution for the chain.
Intuitively, it is clear why  this result holds if we consider a sample path starting at time $-\infty$. When this path reaches time $-T$ it must pick
{\it some} value $x$, and from then on it follows the trajectory from that
value. By construction of $T$,
it arrives at
the same place at time zero no matter what value $x$ is picked at time $-T$,
so the value returned by the algorithm at time zero is the tail end of a sample path that has run for an infinitely long time. Construction of ``every possible sample path'' is usually facilitated by exploiting a monotonicity structure which will allow us to consider only two paths, one from the top of the space and one from the bottom, that will sandwich all possible paths in between.  If the space is unbounded, ``stochastically dominating'' processes can be used to create upper and lower bounds for the process of interest. An $\varepsilon$- perfect algorithm, such as in \cite{mol99}, does not achieve full coupling of all sample paths, but rather gets them arbitrarily close, according to a given measure, in a way that guarantees that they will stay that way. For more details on perfect simulation in general, we refer the reader to \cite{casrob00}, \cite{cortwe01b}, \cite{fostwe97a}, \cite{murgre97}, and \cite{prowil96}.

Our perfect sampler will be based on the Gibbs sampler. For $i=1,2,\ldots, m$, let $x_{-i} = (x_{1}, \ldots, x_{i-1}, x_{i+1}, \ldots, x_{m})$. The standard Gibbs sampler targeting (\ref{e:target}) assumes that one can sample directly from the conditional densities
\beq
\elabel{conditional}
f(x_{i}|x_{-i}) \propto f_{i}(x_{i}) \, I_{\{x_{i-1}< x_{i} < x_{i+1}\}}
\eeq
of the target density (\ref{e:target}) for $i=1,2,\ldots, m$. If we assume that $f_{i}(x)$ has an associated invertible cdf $F_{i}(x)$, we can sample from truncated version of $f_{i}$ in (\ref{e:conditional}) by sampling $U \sim Uniform (F_{i}(x_{i-1}),F_{i}(x_{i+1}))$ and returning $F_{i}^{-1}(U)$.

Although there are many flavors of Gibbs samplers, in its simplest version, a non-perfect forward Gibbs sampler targeting (\ref{e:target}) starts at some arbitrary point $(X_{1}^{(0)}, X_{2}^{(0)}, \ldots, X_{m}^{(0)})$ at time $0$ and moves between time steps $n$ and $n+1$ as described in Algorithm \ref{alg:Gibbs}.

\begin{algorithm}
  \SetAlgoLined
  \caption{The Gibbs Sampler: A Move from Time $n$ to Time $n+1$}
\label{alg:Gibbs}
\begin{algorithmic}
\STATE \textbf{Input:} A starting value $(X_{1}^{(n)}, X_{2}^{(n)}, \ldots, X_{n}^{(n)})$

\vspace{0.1in}
\STATE \textbf{Sample} $X_{1}^{(n+1)} \sim f(x_{1}|X_{2}^{(n)}, \ldots, X_{m}^{(n)}) \propto f_{1}(x_{1}) \, I_{\{-\infty < x_{1} < X_{2}^{(n)}\}}$

\vspace{0.1in}
\For{$j=2$ \TO $m-1$}{
\textbf{Sample} $X_{j}^{(n+1)} \sim f(x_{j}|X_{1}^{(n+1)}, \ldots X_{j-1}^{(n+1)} , X_{j+1}^{(n)} \ldots, X_{m}^{(n)})$

\vspace{0.1in}
\hspace{1.1in}$
\propto f_{n}(x_{n}) \, I_{\{X_{j-1}^{(n+1)}< x_{n} < X_{j+1}^{(n)}\}}$

}

\vspace{0.1in}

\STATE \textbf{Sample} $X_{m}^{(n+1)} \sim f(x_{m}|X_{1}^{(n+1)},  \ldots, X_{m-1}^{(n+1)}) \propto f_{m}(x_{m}) \, I_{\{X_{m-1}^{(n+1)}< x_{m} < \infty\}}$.
\end{algorithmic}
\end{algorithm}

Let $V_{n+1}=(V_{1}^{(n+1)}, V_{2}^{(n+1)}, \ldots, V_{m}^{(n+1)})$ be a vector of  iid $Uniform (0,1)$ random variables. If we define $X_{-i}^{(n)}$ to be $(X_{1}^{(n+1)}, \ldots, X_{i-1}^{(n+1)}, X_{i+1}^{(n)}, \ldots, X_{m}^{(n)})$, we may describe the Gibbs update from time $n$ to time $n+1$ with the stochastic recursive sequence representation given by
\beq
\label{eq:srs}
\begin{array}{lcl}
X_{i}^{(n+1)} &:=& \phi_{i}(X_{-i}^{(n)},V_{i}^{(n+1)})\\
\\
&:=& \left\{
\begin{array}{lcl}
F_{1}^{-1} [F_{1} (X_{2}^{(n)}) \cdot V_{1}^{(n+1)}] &,& i=1\\
\\
F_{i}^{-1} \left[ F_{i} (X_{i-1}^{(n+1)}) + \left( F_{i} (X_{i+1}^{(n)}) - F_{i} (X_{i-1}^{(n+1)}) \right) V_{i}^{(n+1)} \right] &,& i=2,3,\ldots, m-1\\
\\
F_{m}^{-1} \left[ F_{m}(X_{m-1}^{(n+1)}) + \left( 1- F_{m}(X_{m-1}^{(n+1)}) \right) V_{m}^{(n+1)} \right] &,& i=m.
\end{array}
\right.
\end{array}
\eeq

Note that, in all cases, we have that $X_{i}^{(n+1)}$ is an increasing function of $(X_{i-1}^{(n+1)},X_{i+1}^{(n)})$. This stochastic monotonicity will be a key observation to exploit for our backward coupling algorithm. Specifically, if $X_{-i}^{(n)} \preceq Y_{-i}^{(n)}$ where $\preceq$ denotes the product ordering (inequality holding componentwise), we have  $X_{i}^{(n+1)} = \phi_{i}(X_{-i}^{(n)},V_{n+1}) \preceq Y_{i}^{(n+1)} = \phi_{i}(Y_{-i}^{(n)},V_{n+1})$.

\begin{figure}
\caption{Updating paths for backward coupling}
\label{fig:moves}
\centering
\begin{subfigure}{.5\textwidth}
  \centering
\begin{tikzpicture}
  \matrix (m) [matrix of math nodes,row sep=2em,column sep=3em,minimum width=2em]
  {
     X_{4}^{(-n)} & X_{4}^{(-n+1)} \\
     X_{3}^{(-n)} & X_{3}^{(-n+1)} \\
     X_{2}^{(-n)} & X_{2}^{(-n+1)} \\
     X_{1}^{(-n)} & X_{1}^{(-n+1)}\\};
  \path[-stealth]
    (m-1-1) edge node [right, above] {3} (m-2-2)
    (m-2-1) edge node [right, above] {2} (m-3-2)
    (m-3-1) edge node [right, above] {1} (m-4-2)
    (m-2-2) edge node [right] {4} (m-1-2)
    (m-3-2) edge node [right] {3} (m-2-2)
    (m-4-2) edge node [right] {2} (m-3-2);
\end{tikzpicture}
\vspace{0.3in}
  \caption{(a) \,\, one path, one time step}
\end{subfigure}%
\begin{subfigure}{.5\textwidth}
  \centering
  \includegraphics[width=3.5in,height=2.5in]{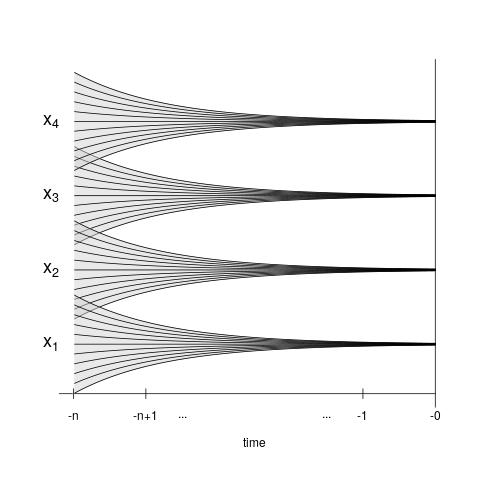}
  \caption{(b) \,\, multiple paths from time $-n$ ideally coalescing to one four-dimensional point}
\end{subfigure}
\end{figure}
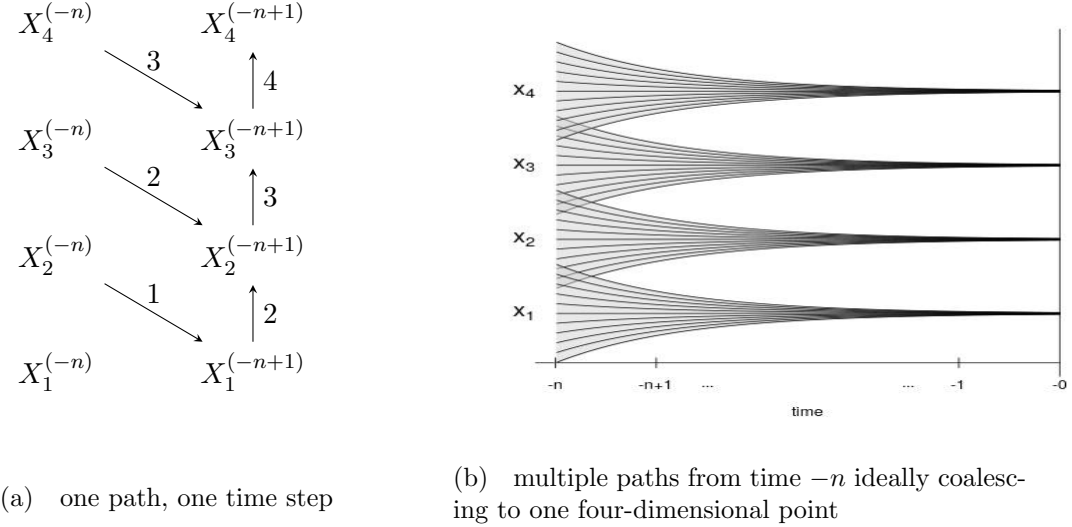

A single time step of a single sample path started at $X_{-n} = (X_{1}^{(-n)},\ldots,X_{m}^{(-n)})$ at some time $-n$ can be obtained through $m$ moves as depicted in Figure \ref{fig:moves}(a) for the case that $m=4$. Here, move numbers are labeled on arrows. We want to achieve a ``coupling'' or ``coalescence'' of paths for all possible values of $X_{i}$ at time $-n$ by time $0$, for each of $i=1,2,3,4$, into four distinct points $X_{1}^{(0)}$, $X_{2}^{(0)}$, $X_{3}^{(0)}$, and $X_{4}^{(0)}$ as depicted in an overly idealized way in Figure \ref{fig:moves}(b). Note that the lower paths, for example depicted in the ``funnel'' for $X_{3}$, are generated by simulating $X_{3}$ from the conditional density $f(x_{3}|x_{1},x_{2},x_{4})= f(x_{3}|x_{2},x_{4})$ using values of $x_{2}$ from the lower paths of the $X_{2}$ funnel and values of $x_{4}$ from the lower paths of the $X_{4}$ funnel. Likewise, upper paths for $X_{3}$ are generated by simulating $X_{3}$ from $f(x_{3}|x_{2},x_{4})$ using upper paths from $X_{2}$ and $X_{4}$.

\subsection{Lower and Upper Processes}

With monotonicity in the model, we will be able to capture the behavior of large collections of sample paths between paths from the ``top'' and ``bottom'' of the space for each of the $m$ components of the chain.  However, if the $f_{i}$ have unbounded support, such points will not exist. For perfect simulation, this typically means that one would have to devise a reversible and stochastically dominating process that could be run backwards from a stationary draw at time $0$.
However, with updates given by (\ref{eq:srs}), it is possible to run upper and lower processes from $\infty$ and $-\infty$, respectively, using the fact that the involved cdfs are increasing and bounded between $0$ and $1$.

Fix a time $-n$ and consider starting sample paths forward to time $0$ from lower and upper points $X_{i,L}^{(-n)}$ and $X_{i,U}^{(-n)}$ for $i=1,2,\ldots, m$. Note that
$$
X_{1}^{(-n)} = F_{1}^{-1}[F_{1}(X_{2}^{(-n-1)}) \cdot V_{1}^{(-n)}] \leq  F_{1}^{-1} [V_{1}^{(-n)}].
$$
So, we define the first component of the upper process started at time $-n$ as
\beq
\elabel{upper1}
X_{1,U}^{(-n)} = F_{1}^{-1} [V_{1}^{(-n)}]
\eeq
which will be above $X_{1}^{(-n)}$ generated by any and all possibilities for  $X_{2}$ at time $-n-1$.

Similarly, for $i=2,3,\ldots, m-1$, we have
$$
\begin{array}{lcl}
X_{i}^{(-n)} &=& F_{i}^{-1} [F_{i}(X_{i-1}^{(-n)}) + (F_{i}(X_{i+1}^{(-n-1)})-F_{i} (X_{i-1}^{(-n)})) \cdot V_{i}^{(-n)}]\\
\\
& \leq & F_{i}^{-1} [F_{i}(X_{i-1,U}^{(-n)}) + (1-F_{i} (X_{i-1,U}^{(-n)})) \cdot V_{i}^{(-n)}].
\end{array}
$$
Thus, we define
\beq
\elabel{uppermid}
X_{i,U}^{(-n)} =  F_{i}^{-1} [F_{i}(X_{i-1,U}^{(-n)}) + (1-F_{i} (X_{i-1,U}^{(-n)})) \cdot V_{i}^{(-n)}]
\eeq
which is an upper bound on $X_{i}^{(-n)}$.

Finally, we have that
\beq
\elabel{upperm}
X_{m,U}^{(-n)} := F_{m}^{-1} [F_{m} (X_{m-1,U}^{(-n)}) + (1-F_{m} (X_{m-1,U}^{(-n)})) \cdot V_{m}^{(-n)}]
\eeq
is an upper bound on 
$$
X_{m}^{(-n)} = F_{m}^{-1} [F_{m} (X_{m-1}^{(-n)}) + (1-F_{m} (X_{m-1}^{(-n)})) \cdot V_{m}^{(-n+1)}]
$$
for all finite $X_{m-1}^{(-n)}$.

In defining this upper process starting at time $n$, we have essentially started at time $-n-1$ with $X_{i,U}^{(-n-1)}=\infty$ for all $i=1,2,\ldots, m$ and used the updates defined in (\ref{eq:srs}).

The lower process is more complicated. Using, for example, $X_{2,L}^{(-n-1)} = -\infty$ would result in an update to the first component of the lower process of
$$
X_{1,L}^{(-n)} = F_{1}^{-1}[F_{1}(X_{2,U}^{(-n-1)}) \cdot V_{1}^{(-n)}] =  F_{1}^{-1} [0] = -\infty.
$$
Indeed, all components of a lower process constructed in this manner would remain ``stuck'' at $-\infty$ for all time. To avoid this, we will instead begin by bounding the largest ($m$th) component of the lower process and working our way down to the smallest component, while taking care to preserve the original update order. Note that
$$
\begin{array}{lcl}
X_{m}^{(-n)} &=& F_{m}^{-1} [F_{m} (X_{m-1}^{(-n)}) + (1-F_{m} (X_{m-1}^{(-n)})) \cdot V_{m}^{(-n)}]\\
\\
& \geq & F_{m}^{-1} [F_{m} (-\infty) + (1-F_{m} (-\infty)) \cdot V_{m}^{(-n)}] = F_{m}^{-1} [ V_{m}^{(-n)}] =: X_{m,L}^{(-n)}
\end{array}
$$
is a lower bound for $X_{m}^{(-n)}$ for all $n$. Similarly,
$$
\begin{array}{lcl}
X_{m-1}^{(-n)} &=& F_{m-1}^{-1} [F_{m-1}(X_{m-2}^{(-n)}) + (F_{m-1}(X_{m}^{(-n-1)})-F_{m-1} (X_{m-1}^{(-n)})) \cdot V_{i}^{(-n)}]\\
\\
&\geq & F_{m-1}^{-1}[F_{m-1}(X_{m}^{(-n-1)}) \cdot V_{i}^{(-n)}] \geq F_{m-1}^{-1}[F_{m-1}(X_{m,L}^{(-n-1)}) \cdot V_{i}^{(-n)}] =: X_{m-1,L}^{(-n)}
\end{array}
$$
where $X_{m,L}^{(-n-1)} = F_{m}^{-1} [ V_{m}^{(-n-1)}]$. As we move through the components, in decreasing order, to define the lower bounds  $X_{i,L}^{(-n)}$ for $i=m-2,m-3,\ldots, 1$, we note two things. The first is that, the lower bound for the first component at time $-n$ depends on the lower bound for the second component at time $-n-1$ which, in turn, depends on the lower bound for the third component at time $-n-2$. Continuing in this way, we see that the lower bound on the first component at time $-n$ depends, ultimately, on the lower bound for the $m$th component at time $-n-m+1$, which is $X_{m,L}^{(-n-m+1)}$. Thus, in order to attempt backward coupling starting from time $-n$ to a common value at time $0$, we need to generate random vectors $V_{-n}, V_{-n-1}, \ldots, V_{-n-m+1}$ in addition to the usual $V_{-n+1}, V_{-n+2}, \ldots, V_{0}$. The second thing to note is that these lower bounds for our order statistics are not necessarily in order themselves. We could adjust the lower bounds as we go, for example replacing $X_{m,L}^{(-n)}$ with $\max(X_{m-1,L}^{(-n)},X_{m,L}^{(-n)})$, or we could simply use $X_{1,L}^{(-n)}$ as a lower bound for $X_{i}^{(-n)}$ for all $i=1,2,\ldots, m$. In this paper we chose the later approach, for simplicity, even though it will increase the backward coupling time. $X_{1,L}^{(-n)}$ is computed as follows.

\begin{algorithm}
  \SetAlgoLined
  \caption{A Starting Point at Time $-n$ for the Lower Bounding Process}
\label{alg:lower}
\begin{algorithmic}
\STATE \textbf{Compute:} $X_{m,L}^{(-n-m+1)} = F_{m}^{-1}[V_{m}^{(-n-m+1)}]$.

\vspace{0.1in}

\For{$i=m-1$ \TO $2$}{
\textbf{Compute:} $X_{i,L}^{(-n-i+1)} := F_{i}^{-1}[F_{i}(X_{i+1,L}^{(-n-i)}) \cdot V_{i}^{(-n-i+1)}]$.
}

\vspace{0.1in}
\STATE \textbf{Set:} $X_{1,L}^{(-n)} = F_{1}^{-1}[F_{1}(X_{2,L}^{(-n-1)}) \cdot V_{1}^{(-n)}]$.

\end{algorithmic}
\end{algorithm}

%\begin{itemize}
%\item Compute 
%\beq
%\elabel{lowerm}
%X_{m,L}^{(-n-m+1)} = F_{m}^{-1}[V_{m}^{(-n-m+1)}].
%\eeq
%\item For $i=m-1,m-2,\ldots, 2$, compute
%\beq
%\elabel{lowermid}
%X_{i,L}^{(-n-i+1)} := F_{i}^{-1}[F_{i}(X_{i+1,L}^{(-n-i)}) \cdot V_{i}^{(-n-i+1)}].
%\eeq
%\item Set 
%\beq
%\elabel{lower1}
%X_{1,L}^{(-n)} = F_{1}^{-1}[F_{1}(X_{2,L}^{(-n-1)}) \cdot V_{1}^{(-n)}].
%\eeq
%
%\end{itemize}

\subsection{The Algorithm}
Now that we have starting values for upper and lower processes bounding each component of the sample path of interest for some time $-n$, we evolve the paths forward to time $0$ with a Gibbs sampler. If $\sum_{i=1}^{m} (X_{i,L}^{(0)}-X_{i,U}^{(0)})^{2} < \varepsilon$, we output
$$
\left( \frac{X_{1,L}^{(0)} + X_{1,U}^{(0)}}{2}, \frac{X_{1,L}^{(1)} + X_{1,U}^{(1)}}{2}, \ldots, \frac{X_{m,L}^{(0)} + X_{m,U}^{(0)}}{2} \right)
$$  
as an $\varepsilon$-perfect draw from (\ref{e:target}).
If the stopping criterion is not met, we start further back in time while making sure to reuse any previously generated $V_{i}^{(-k)}$ on subsequent passes over time steps $-k$ closer to $0$. More specific  details are given in Algorithm \ref{thealgorithm}.

\begin{algorithm}
  \SetAlgoLined
 \caption{Perfect Gibbs Sampler for OCNID Order Statistics (single draw)}
\label{thealgorithm}
\begin{algorithmic}

  \STATE \textbf{Initialization:} Set $n=1$ and set an error tolerance $\varepsilon>0$. Generate and store $m$-dimensional independent vectors $V_{-m},V_{-m+1}, \ldots, V_{0}$ where each is populated with iid $Uniform(0,1)$ random variates. Set $exit=FALSE$.\\
\vspace{0.1in}
  \While{$exit =$ FALSE}{
    \textbf{Compute:} $X_{1,U}^{(-n)}, X_{2,U}^{(-n)}, \ldots, X_{m,U}^{(-n)}$ using (\ref{e:upper1}), (\ref{e:uppermid}), and (\ref{e:upperm}).\\
\textbf{Compute:} $X_{1,L}^{(-n)}$ using Algorithm \ref{alg:lower}. Set $X_{i,L}^{(-n)} = X_{1,L}^{(-n)}$ for $i=2,3,\ldots,m$.
\\
\vspace{0.1in}
\For{$k=n-1$ down to $0$}{

\vspace{0.1in}
\textbf{Compute:} 
$$
\begin{array}{lcl}
X_{1,L}^{(-k)} &=& F_{1}^{-1}[F_{1}(X_{2,L}^{(-k-1)}) \cdot V_{1}^{(-k)}] \,\,\,\,\, \mbox{and}\\
\\
X_{1,U}^{(-k)} &=& F_{1}^{-1}[F_{1}(X_{2,U}^{(-k-1)}) \cdot V_{1}^{(-k)}]. 
\end{array}
$$\\

\For{$i=2$ down to $m-1$}{

\vspace{0.1in}
\textbf{Compute:}
$$
\begin{array}{lcl}
X_{i,L}^{(-k)} &=& F_{i}^{-1}[F_{i}(X_{i-1,L}^{(-k)})+(F_{i}(X_{i+1,L}^{(-k-1)})-F_{i}(X_{i-1,L}^{(-k)})) \cdot V_{i}^{(-k)}]  \,\,\,\,\, \mbox{and}\\
\\
X_{i,U}^{(-k)} &=& F_{i}^{-1}[F_{i}(X_{i-1,U}^{(-k)})+(F_{i}(X_{i+1,U}^{(-k-1)})-F_{i}(X_{i-1,U}^{(-k)})) \cdot V_{i}^{(-k)}]
\end{array}
$$
}

\vspace{0.1in}
\textbf{Compute:}
$$
\begin{array}{lcl}
X_{m,L}^{(-k)} = F_{m}^{-1} [F_{m}(X_{m-1,L}^{(-k)}) + (1-F_{m}(X_{m-1,L}^{(-k)}))\cdot V_{m}^{(-k)}] \,\,\,\,\, \mbox{and}\\
\\
X_{m,U}^{(-k)} = F_{m}^{-1} [F_{m}(X_{m-1,U}^{(-k)}) + (1-F_{m}(X_{m-1,U}^{(-k)}))\cdot V_{m}^{(-k)}].
\end{array}
$$

}
    \eIf{$\sum_{i=1}^{m} (X_{i,L}^{(0)}-X_{i,U}^{(0)})^{2} < \varepsilon$}{
      
      \vspace{0.1in}
      \textbf{Output:} 
$$
\left( \frac{X_{1,L}^{(0)} + X_{1,U}^{(0)}}{2}, \frac{X_{1,L}^{(1)} + X_{1,U}^{(1)}}{2}, \ldots, \frac{X_{m,L}^{(0)} + X_{m,U}^{(0)}}{2} \right).
$$
      Set $exit = TRUE$\\
      }{
      Let $n=n+1$. Generate and store an $m$-dimensional vector $V_{-n-m+1}$ of iid $Uniform(0,1)$ random variates. 
    }
  }

\end{algorithmic}
\end{algorithm}

\section{Examples}
In this Section, we give several examples and simulation results for the perfect Gibbs OCNID sampler which show the performance of the sampler for scale parameter distributions, shape parameter distributions, heavy-tailed distributions, and a case  where the $f_{i}$ in (\ref{e:target}) come from completely different families.

\subsection{Exponential Distributions}
\slabel{expsec}

\begin{figure*}
 \caption[]
        {\small Simulation results for the exponential distribution with $m=4$, $\theta=(8,6,4,2)$ and $\varepsilon= 0.0001$ with exact marginal densities superimposed} 
        \centering
        \begin{subfigure}[b]{0.475\textwidth}
            \centering
            \includegraphics[width=\textwidth]{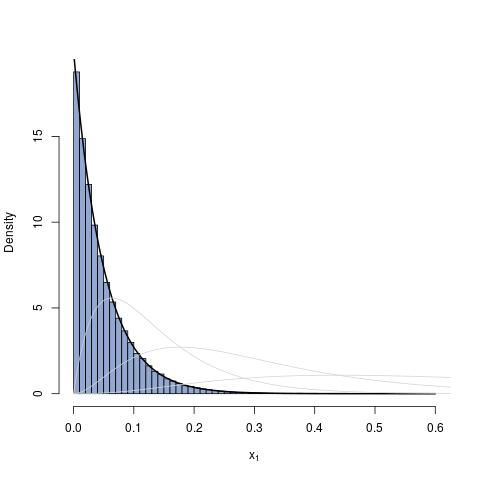}
            \caption[Network2]%
            {{\small (a) $X_{1}$}}    
        \end{subfigure}
        \hfill
        \begin{subfigure}[b]{0.475\textwidth}  
            \centering 
            \includegraphics[width=\textwidth]{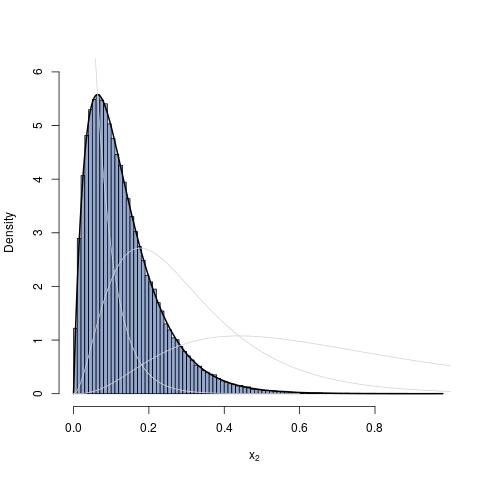}
            \caption[]%
            {{\small (b) $X_{2}$}}    
        \end{subfigure}
        \vskip\baselineskip
        \begin{subfigure}[b]{0.475\textwidth}   
            \centering 
            \includegraphics[width=\textwidth]{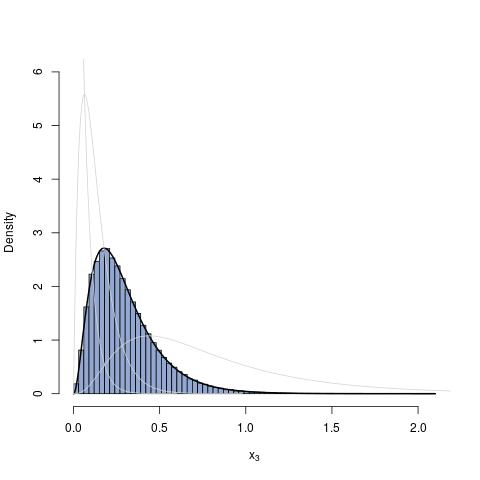}
            \caption[]%
            {{\small (c) $X_{3}$}}    
        \end{subfigure}
        \quad
        \begin{subfigure}[b]{0.475\textwidth}   
            \centering 
            \includegraphics[width=\textwidth]{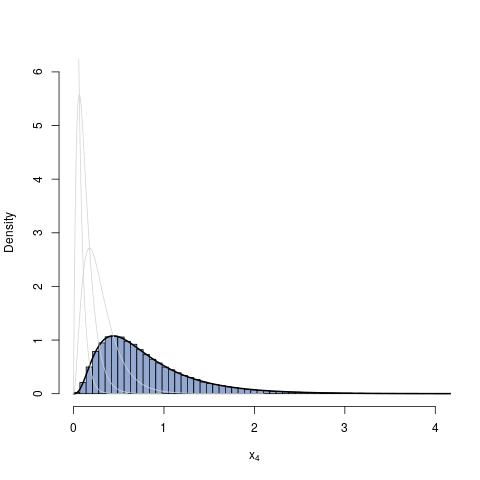}
            \caption[]%
            {{\small (d) $X_{4}$}}    
        \end{subfigure}
       
        \label{fig:exp}
    \end{figure*}

Here, the $f_{i}$ in (\ref{e:target}) are exponential distributions with various parameters. Taking
$$
f_{i}(x) = \theta_{i} e^{-\theta_{i} x} \, I_{(0,\infty)}(x),
$$
we simulated 100,000 OCNID order statistics with $m = 4$, $\theta = (8,6,4,2)$ and $\varepsilon = 0.0001$. Histograms of the results are shown in Figure \ref{fig:exp}. The superimposed curves are the exact marginal pdfs which are tractable for this example. The mean BCT was $7.4$ with a minimum of $1$ and a maximum of $17$. In Table \ref{expbct} we show the mean backward coupling times for various $m$ and inverse scale parameters $\theta_{i}$ chosen to be smaller, larger, and with more range. For this example, the $\theta_{i}$ are chosen in decreasing order, though this is not a requirement for our algorithm. Using increasing or other orderings for $\theta$ will increase the backward coupling time, though there is empirical evidence that it will be of the same order of magnitude. For example, using $\theta=(2,4,6,8)$ instead of $\theta=(8,6,4,2)$ increased the mean BCT in $100,000$ draws from $7.4$ to $11.2$. 

\begin{table}[h]
\caption{Mean backward coupling times for the exponential, Weibull, Cauchy, and Pareto  distributions}
\label{expbct}
\centering
\begin{tabular}{cccccc}
\hline
 $m$ & $\theta$ &  \multicolumn{4}{c}{\underline{mean BCT}}\\
 & & exponential & Weibull & Cauchy & Pareto\\
\hline
\multirow{ 4}{*}{4} & $(8,6,4,1)$ & $8.4$ & $5.6$ & $12.1$ & $9.1$\\
 & $(20,5,2,1)$ & $6.9$ & $4.8$ & $11.9$ & $7.8$\\
 & $(1.2,0.8,0.2,0.05)$ & $8.1$ & $5.6$ & $12.2$ & $20.7$\\
 & $(50,30,20,10)$ & $6.2$ & $4.3$ & $11.0$ & $6.2$ \\
\hline
\multirow{ 4}{*}{8} & $(20,14,10,8,6,5,4,2)$ & $24.1$ & $11.3$ & $45.8$ & $26.5$\\
 & $(50,12,10,6,4,2,0.5,0.1)$ & $16.0$ & $9.0$ & $32.8$ & $38.6$\\
 & $(5,2,1.9,1.2,0.6,0.4,0.2,0.1)$ & $23.3$ & $10.7$ & $42.0$ & $47.2$\\
 & $(100,70,50,30,20,10,5,1)$ & $13.4$ & $7.7$ & $26.9$ & $14.2$\\
\hline
\multirow{ 4}{*}{12} & $(20,18,14,12,10,8,7,6,5,4,2,1)$ & $46.0$ & $19.0$ & $87.4$ & $78.7$\\
& $(70,50,14,12,10,8,7,6,5,4,0.2,0.11)$ & $36.6$ & $16.4$ & $72.3$ & $80.1$\\
& $(4.5,4,3.5,3.2,2,1.9,1.2,0.6,0.4,0.3,0.2,0.1)$ & $44.3$ & $18.7$ & $80.5$ & $91.7$\\
& $(100,90,80,70,50,40,30,20,10,8,5,1)$ & $26.2$ & $12.2$ & $59.1$ & $32.3$\\
\end{tabular}
\end{table}

\subsection{Weibull Distributions}

\begin{figure*}
 \caption[]
        {\small Simulation results for the Weibull distribution with $m=4$, $\alpha=3$, $\theta=(8,6,4,2)$ and $\varepsilon= 0.0001$ with exact marginal densities superimposed} 
        \centering
        \begin{subfigure}[b]{0.475\textwidth}
            \centering
            \includegraphics[width=\textwidth]{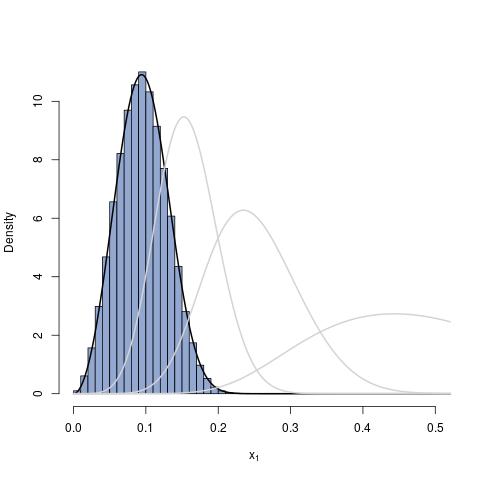}
            \caption[Network2]%
            {{\small (a) $X_{1}$}}    
        \end{subfigure}
        \hfill
        \begin{subfigure}[b]{0.475\textwidth}  
            \centering 
            \includegraphics[width=\textwidth]{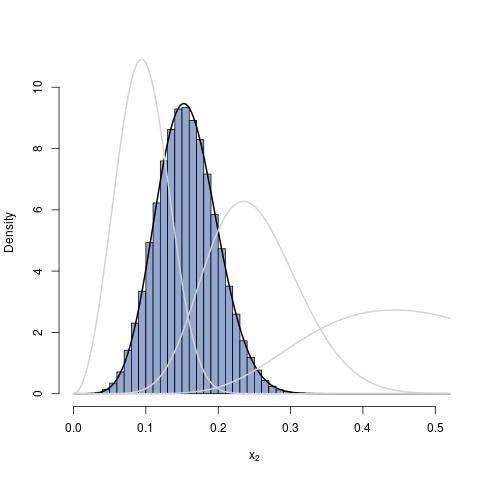}
            \caption[]%
            {{\small (b) $X_{2}$}}    
        \end{subfigure}
        \vskip\baselineskip
        \begin{subfigure}[b]{0.475\textwidth}   
            \centering 
            \includegraphics[width=\textwidth]{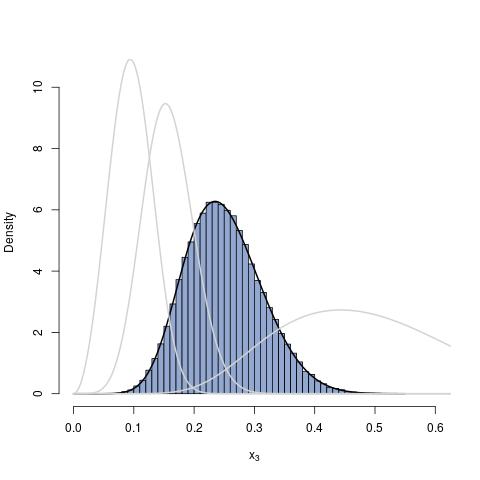}
            \caption[]%
            {{\small (c) $X_{3}$}}    
        \end{subfigure}
        \quad
        \begin{subfigure}[b]{0.475\textwidth}   
            \centering 
            \includegraphics[width=\textwidth]{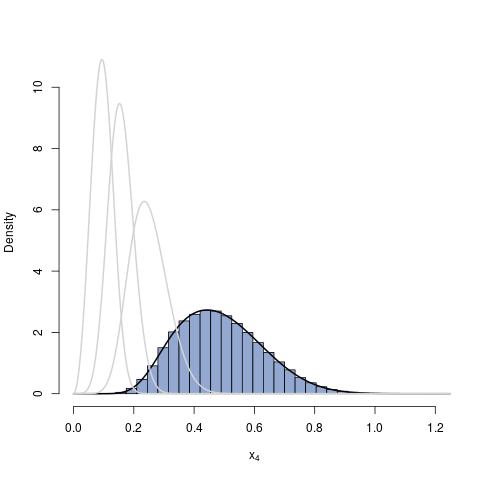}
            \caption[]%
            {{\small (d) $X_{4}$}}    
        \end{subfigure}
       
        \label{fig:weibull}
    \end{figure*}

Here, we used
$$
f_{i}(x) = \alpha \theta_{i}^{\alpha} x^{\alpha-1} e^{-(\theta_{i} x)^{\alpha}} \, I_{(0,\infty)}(x)
$$
with $\alpha=3$.
We simulated 100,000 OCNID order statistics with $m = 4$, $\theta = (8,6,4,2)$ and $\varepsilon = 0.0001$. Histograms of the results are shown in Figure \ref{fig:weibull}. Again, we were able to compute the exact marginal pdfs which are superimposed The mean BCT was $4.6$ with a minimum of $1$ and a maximum of $9$. Other backward coupling times for various $m$ and parameter choices are included in Table \ref{expbct}.

%\begin{table}[h]
%\caption{Mean backward coupling times for the Weibull %distribution}
%\label{weibullbct}
%\centering
%\begin{tabular}{ccc}
%\hline
% $m$ & $\theta$ & mean BCT \\
%\hline
%\multirow{ 4}{*}{4} & $(8,6,4,1)$ & $5.6$\\
% & $(20,5,2,1)$ & $4.8$\\
% & $(1.2,0.8,0.2,0.05)$ & $5.6$\\
% & $(50,30,20,10)$ & $4.3$\\
%\hline
%\multirow{ 4}{*}{8} & $(20,14,10,8,6,5,4,2)$ & $11.3$\\
% & $(50,12,10,6,4,2,0.5,0.1)$ & $9.0$\\
% & $(5,2,1.9,1.2,0.6,0.4,0.2,0.1)$ & $10.7$\\
% & $(100,70,50,30,20,10,5,1)$ & 7.7\\
%\hline
%\multirow{ 4}{*}{12} & $(20,18,14,12,10,8,7,6,5,4,2,1)$ & %$19.0$\\
% & $(70,50,14,12,10,8,7,6,5,4,0.2,0.11) & $16.4$\\
% & $(4.5,4,3.5,3.2,2,1.9,1.2,0.6,0.4,0.3,0.2,0.1)$ & %$18.7$\\
% & $(100,90,80,70,50,40,30,20,10,8,5,1)$ & $12.2$\\
%\end{tabular}
%\end{table}

\subsection{Cauchy Distributions}

\begin{figure*}
 \caption[]
        {\small Simulation results for the Cauchy distribution with $m=4$, $\theta=(8,6,4,2)$ and $\varepsilon= 0.0001$} 
        \centering
        \begin{subfigure}[b]{0.475\textwidth}
            \centering
            \includegraphics[width=\textwidth]{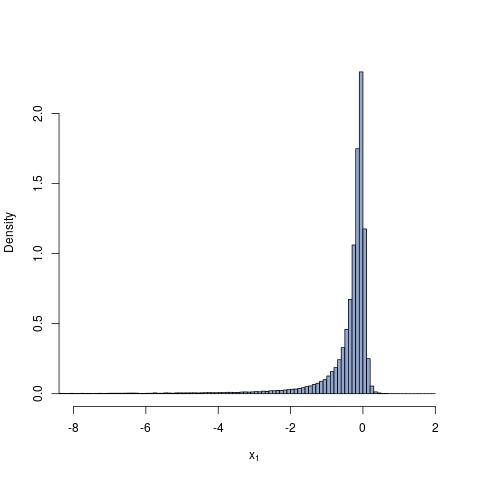}
            \caption[Network2]%
            {{\small (a) $X_{1}$}}    
        \end{subfigure}
        \hfill
        \begin{subfigure}[b]{0.475\textwidth}  
            \centering 
            \includegraphics[width=\textwidth]{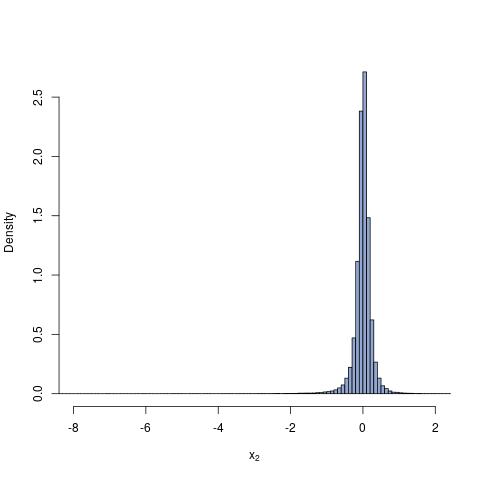}
            \caption[]%
            {{\small (b) $X_{2}$}}    
        \end{subfigure}
        \vskip\baselineskip
        \begin{subfigure}[b]{0.475\textwidth}   
            \centering 
            \includegraphics[width=\textwidth]{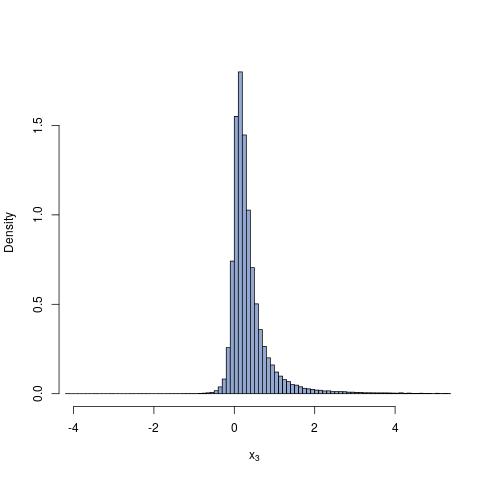}
            \caption[]%
            {{\small (c) $X_{3}$}}    
        \end{subfigure}
        \quad
        \begin{subfigure}[b]{0.475\textwidth}   
            \centering 
            \includegraphics[width=\textwidth]{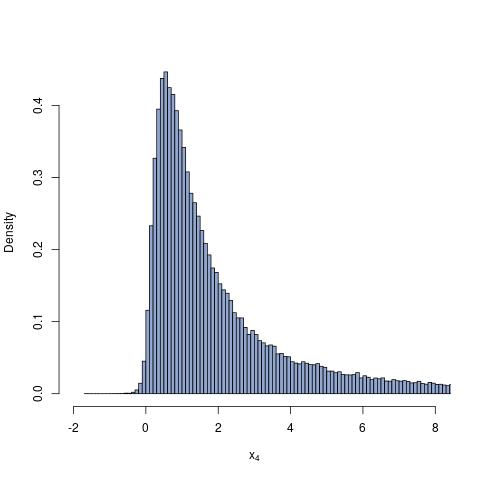}
            \caption[]%
            {{\small (d) $X_{4}$}}    
        \end{subfigure}
       
        \label{fig:cauchy}
    \end{figure*}

As a heavy-tailed example where computation of the exact marginal pdfs is intractable, we
we simulated 100,000 OCNID order statistics using  the Cauchy pdfs 
$$
f_{i}(x) = \frac{\theta_{i}}{\pi} \frac{1}{1+(\theta_{i} x)^{2}}
$$
to define $f_{i}(x) = \theta_{i} f_{0}(\theta_{i} x)$.
We again used $m = 4$, $\theta = (8,6,4,2)$ and $\varepsilon = 0.0001$. Histograms of the results are shown in Figure \ref{fig:cauchy}. The mean BCT was $11.1$ with a minimum of $4$ and a maximum of $37$. Other backward coupling times are given in Table \ref{expbct}.

%\begin{table}[h]
%\caption{Mean backward coupling times for the Cauchy distribution}
%\label{cauchybct}
%\centering
%\begin{tabular}{ccc}
%\hline
% $m$ & $\theta$ & mean BCT \\
%\hline
%\multirow{ 4}{*}{4} & $(8,6,4,1)$ & $12.1$\\
% & $(20,5,2,1)$ & $11.9$\\
% & $(1.2,0.8,0.2,0.05)$ & $12.2$\\
% & $(50,30,20,10)$ & $11.0$\\
%\hline
%\multirow{ 4}{*}{8} & $(20,14,10,8,6,5,4,2)$ & 45.8\\
% & $(50,12,10,6,4,2,0.5,0.1)$ & $32.8$\\
% & $(5,2,1.9,1.2,0.6,0.4,0.2,0.1)$ & $42.0$\\
% & $(100,70,50,30,20,10,5,1)$ & $26.9$\\
%\hline
%\multirow{ 4}{*}{12} & $(20,18,14,12,10,8,7,6,5,4,2,1)$ & $87.4$\\
%& $(70,50,14,12,10,8,7,6,5,4,0.2,0.11) & $72.3$\\
%& $(4.5,4,3.5,3.2,2,1.9,1.2,0.6,0.4,0.3,0.2,0.1)$ & $80.5$\\
% & $(100,90,80,70,50,40,30,20,10,8,5,1)$ & $59.1$\\
%\end{tabular}
%\end{table}

\subsection{Pareto Distributions}

As a heavy-tailed example with a shape, as opposed to scale,   parameter,
we simulated 100,000 OCNID order statistics using the Pareto pdfs
$$
f_{i}(x) = \frac{\theta_{i}}{(1+x)^{\theta_{i}+1}} \, I_{(0,\infty)}.
$$
We again used $m = 4$, $\theta = (8,6,4,2)$ and $\varepsilon = 0.0001$. The mean BCT in 100,000 draws was $9.1$ with a minimum of $2$ and a maximum of $22$. Other backward coupling times are given in Table \ref{expbct} though we have omitted histograms for this example for the sake of brevity. We can see expected spikes in backward coupling times in the cases where some $\theta_{i} \leq 1$ and the mean of the corresponding marginal distribution is infinite.

\subsection{Exponential, Weibull, and a ``Folded" Cauchy Distribution}

\begin{figure*}
 \caption[]
        {\small Simulation results with $m=3$, different base distributions given by  (\ref{e:weibullmiddle}), and $\varepsilon= 0.0001$} 
        \centering
        \begin{subfigure}[b]{0.32\textwidth}
            \centering
            \includegraphics[width=\textwidth]{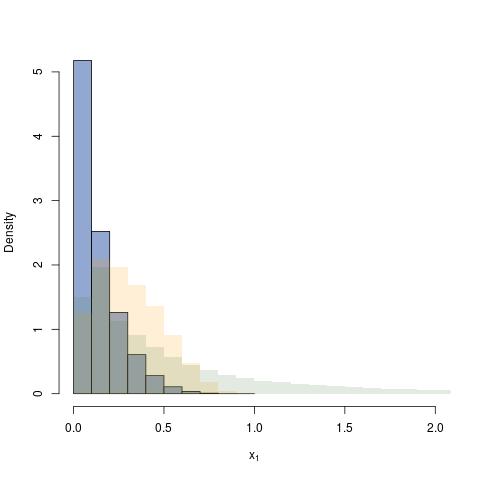}
            \caption[Network2]%
            {{\small (a) $X_{1}$}}    
            \label{fig:mean and std of net14}
        \end{subfigure}
        \begin{subfigure}[b]{0.32\textwidth}  
            \centering 
            \includegraphics[width=\textwidth]{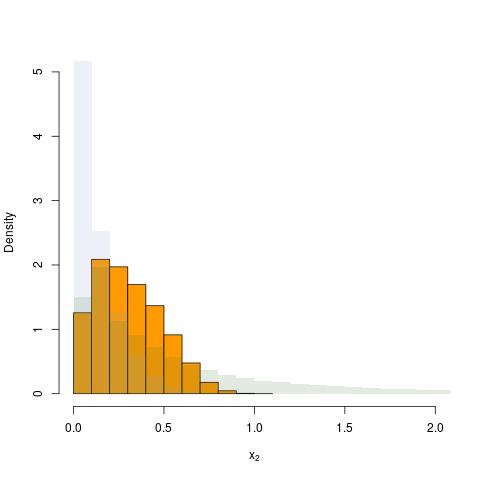}
            \caption[]%
            {{\small (b) $X_{2}$}}    
            \label{fig:mean and std of net24}
        \end{subfigure}
        \begin{subfigure}[b]{0.32\textwidth}   
            \centering 
            \includegraphics[width=\textwidth]{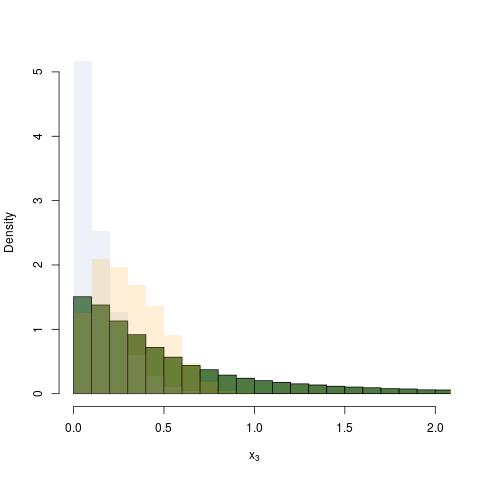}
            \caption[]%
            {{\small (c) $X_{3}$}}    
            \label{fig:mean and std of net34}
        \end{subfigure}
        \label{fig:exp_weib_cauchy}
    \end{figure*}

As an example where the $f_{i}$  come from different families of distributions, we simulated 100,000 values from (\ref{e:target}) using
$$
f_{1}(x) = 2 e^{-2x} \, I_{(0,\infty)}(x),
$$
\beq
\elabel{weibullmiddle}
f_{2}(x) = 3 (2x)^{2} e^{-(2x)^{3}} \, I_{(0,\infty)}(x),
\eeq
and
$$
f_{3}(x) = \frac{2}{\pi} \frac{1}{1+(2x)^{2}} \, I_{(0,\infty)}(x).
$$
The average BCT in $100,000$ draws was $5.2$ (with $\varepsilon = 0.0001$) with a minimum of $1$ and a maximum of $23$. The resulting marginal distributions for $X_{1}$, $X_{2}$, and $X_{3}$ are shown in Figure \ref{fig:exp_weib_cauchy}.

\subsection{Bayesian Principal Components Analysis}
\slabel{bpca}

%Principal component analysis (PCA) \cite{jolliffe1986}  is a widely used method for transforming a data set in order to convert a collection of possibly correlated variables into a collection of linearly uncorrelated variables. The uncorrelated variables, known as {\emph{principal components}}, form a basis for reconstructing the data set. The principal components can be ranked according to their importance in explaining variability in the data and  often lower ranked components can be dropped, creating a smaller basis that can recreate the original data set with varying degrees of accuracy. In 1997, Tipping and Bishop \cite{tippingbishop1999} introduced a density-based variant of PCA, known as {\emph{probabilistic principal component analysis}} (PPCA) which allows for many extensions of the analysis technique to situations that are better suited for data generated from probabilistic models. In 1999 Bishop \cite{bishop99} first used the probabilistic framework of PPCA in order to give a fully Bayesian treatment to estimation of model parameters. Most importantly, {\emph{Bayesian principal component analysis}} (BPCA) allows one to automatically choose the effective dimensionality by maximizing a marginal likelihood. As with most Bayesian estimation and model selection problems, one is required to employ Markov chain Monte Carlo (MCMC) methods in order to sample from posterior distributions for the purpose of inference.

We wish to relate a $d$-dimensional observed vector $\vec{y}$ to a $q$-dimensional vector $\vec{x}$ of latent variables with $1 \leq q< d$. The most common model is the linear model given by
$$
\vec{y} = W \vec{x} + \vec{\mu} + \vec{\varepsilon} 
$$ 
where $W$ is a $d \times q$ matrix relating $\vec{x}$ to $\vec{y}$, $\vec{\mu}$ is a $d \times 1$ vector of constants and $\vec{\varepsilon}$ is a $d \times 1$ vector of random variables. It is typically assumed (e.g. \cite{tippingbishop1999}) that $\vec{x} \sim N(\vec{0},I_{q})$ where $I_{q}$ is the $q \times q$ identity matrix and that $\vec{\varepsilon} \sim N(\vec{0},\sigma^{2} I_{d})$ is isotropic noise. We then have
$$
\vec{y}|\vec{x} \sim N(W \vec{x}+ \mu,\sigma^{2} I_{d})
$$
and it is routine to show then that
$$
\vec{y} \sim N(\vec{\mu},V)
$$
where $V = WW^{T}+ \sigma^{2} I_{d}$.

Given $N$ observations $\vec{y}_{1}, \vec{y}_{2}, \ldots, \vec{y}_{N}$ from this model, the joint pdf is
$$
f(\vec{y}_{1}, \vec{y}_{2}, \ldots, \vec{y}_{N}|\vec{\mu},W,\sigma^{2}) = (2 \pi)^{-dN/2} |V|^{-N/2} \, \exp \left[ -\frac{1}{2} \sum_{n=1}^{N} (\vec{y}_{n}-\vec{\mu})^{T} V^{-1} (\vec{y}_{n}-\vec{\mu}) \right]
$$
which leads to a likelihood given by
\beq
\elabel{likelihood}
L_{q}(\vec{\mu},W,\sigma^{2}) = |V|^{-N/2} \, \exp \left[ -\frac{N}{2} \, tr(V^{-1} S)\right]
\eeq
where $S=\frac{1}{N} \sum_{n=1}^{N} (\vec{y}_{n} -\vec{\mu})(\vec{y}_{n}-\vec{\mu})^{T}$.

Following Minka \cite{minka2001}, we reparameterize the model by expressing $W$ as $W=U_{d,q}(\Lambda_{q}-\sigma^{2}I_{q})^{1/2}$ and  extending the $d \times q$ matrix $U_{d,q}$ of principal eigenvectors of $S$ to a $d \times d$ matrix $U=(U_{d,q},U_{d-q})$ with an orthogonal $(d-q)\times (d-q)$ matrix $U_{d-q}$ so that $U$ is the matrix whose columns are all eigenvectors of $S$ ordered by corresponding descending eigenvalues.

Define $\Lambda$ to be the $d \times d$ matrix
$$
\Lambda = \left[\begin{array}{cc} \Lambda_{q}-\sigma^{2} I_{q} & 0 \\ 0 & 0 \end{array} \right].
$$
Then we have $U \Lambda U^{T}=WW^{T}$ and 
$$
V = WW^{T} =\sigma^{2} I_{d} = U \left[ \begin{array}{cc} \Lambda_{q} & 0\\ 0 & \sigma^{2} I_{d-q} \end{array} \right] U^{T}.
$$
Writing the eigendecompositon of $\widehat{S}$ as $S=AGA^{T}$ where $G=diag(g_{1}, g_{2}, \ldots, g_{d})$ is the diagonal matrix of descending eigenvalues of $\widehat{S}$, we can rewrite the trace in the exponent of (\ref{e:likelihood}) as
$$
tr(V^{-1}S) = tr(U B U^{T} A G A^{T})
$$
where 
$$
B =  \left[ \begin{array}{cc} \Lambda_{q} & 0\\ 0 & \sigma^{2} I_{d-q} \end{array} \right]^{-1}.
$$
Note that $A$ is the MLE of $U$. Following  \cite{zhang2004b}, we simplify things by setting $U=A$. We then have that
$$
tr(V^{-1}S) = tr(A B A^{T} A G A^{T}) = tr( A^{T} A B A^{T} A G) = tr(BG) = \sum_{i=1}^{q} \frac{g_{i}}{\lambda_{i}} + \sum_{i=q+1}^{d} \frac{g_{i}}{\sigma^{2}}
$$
and, for fixed $q$, (\ref{e:likelihood}) can be rewritten as
\beq
\elabel{likelihood2}
L_{q}(\lambda_{1}, \lambda_{2}, \ldots, \lambda_{q},\sigma^{2}) = \left( \prod_{i=1}^{q} \lambda_{i} \right)^{-N/2} \left( \sigma^{2} \right)^{-N(d-q)/2} \exp \left[ -\frac{N}{2} \sum_{i=1}^{q} \frac{g_{i}}{\lambda_{i}} \right] \cdot \exp \left[ -\frac{N}{2 \sigma^{2}} c(q) \right]
\eeq
where $c(q):=\sum_{i=q+1}^{d} g_{i}$.

\subsection*{The Priors}

We will a priori take $q$ to be uniformly distributed on $\{1,2, \ldots, d-1\}$, though other choices of priors can easily be used with our algorithm. It is common in the literature to require that $\lambda_{1}>\lambda_{2} > \cdots > \lambda_{q} > \sigma^{2}$ for identifiability and for $(\lambda_{1}, \lambda_{2}, \ldots, \lambda_{q},\sigma^{2})$ to have the distribution of the order statistics for $q+1$ iid inverse gamma ($IG$) random variables with hyperparameters $\alpha$ and $\beta$ which we will assume are fixed. To be clear, we are using $\beta$ as an inverse scale parameter. Specifically, if $X \sim IG(\alpha,\beta)$ the pdf is 
$$
IG(x;\alpha,\beta) = \frac{1}{\Gamma(\alpha)} \beta^{\alpha} x^{-(\alpha+1)} e^{-\beta/x} \, I_{(0,\infty)}(x).
$$

Define $\theta_{q} = (\lambda_{1}, \lambda_{2}, \ldots, \lambda_{q},\sigma^{2})$. The prior on $q$ and $\theta_{q}$ is
$$
\begin{array}{lcl}
\pi(q,\theta_{q}) &=& \pi(\theta_{q}|q) \, \pi(q)\\
\\
&=& (q+1)! \cdot \left[ \prod_{i=1}^{q} IG (\lambda_{i}; \alpha,\beta) \right] \cdot IG(\sigma^{2};\alpha,\beta) \cdot I_{ \{ \lambda_{1}> \lambda_{2}> \cdots > \lambda_{q} > \sigma^{2} \} } \\
\\
&&\cdot \frac{1}{d-1} \, I_{  \{ 1,2,\ldots, d-1\} }(q).
\end{array}
$$

\subsection*{The Posterior Target}
Let $D := \{\vec{y}_{1}, \vec{y}_{2}, \ldots, \vec{y}_{n}\}$. The posterior distribution is then
\beq
\elabel{posterior}
\pi(q,\theta_{q}|D)  \propto L_{q}(\lambda_{1}, \lambda_{2}, \ldots, \lambda_{q},\sigma^{2}) \cdot \pi(\theta_{q}|q) \, \pi(q)
\eeq
This is the target distribution that we wish to sample from. Note that $\lambda_{1}, \lambda_{2}, \ldots, \lambda_{q}, \sigma^{2}$ are OCNID order statistics. In particular, we have the marginal distributions
$$
\lambda_{i} \sim IG \left( \frac{N}{2}+ \alpha, \frac{N g_{i}}{2}+ \beta\right),
$$
for $i=1,2,\ldots, q$, 
and
$$
\sigma^{2} \sim IG \left(  \frac{N(d-q)}{2} + \alpha, \frac{N c(q)}{2} + \beta \right).
$$
along with the restrictions $\lambda_{1}>\lambda_{2}> \cdots, \lambda_{q}> \sigma^{2}$.

In \cite{zhang2004b}, the authors sample from (\ref{e:posterior}) using a $(q+1)-$dimensional Gibbs  sampler within a reversible jump MCMC algorithm to move $q$ through the set $\{1,2,\ldots, d-1\}$. The $q+1$ parameters each have individual, and non-perfect, Gibbs moves. We have chosen to forgo the reversible jump aspect and, instead, to repeatedly sample $\theta_{q}$ for each fixed value of $q$ in order to estimate a maximum a posteriori (MAP) estimator which can be used for model selection. In particular, we will compare the evidence for each subspace dimensionality, using Laplace's method and the Bayesian Information Criterion (BIC) as in \cite{kazianka2016} and \cite{minka2001}.

We generated a data set of 100 points from an 8-dimensional Gaussian distribution with variances $10$, $8$, $6$, $4$, $2$, $0.5$, $0.5$, and $0.5$.  The observed sample variances were $9.5682, 7.8954, 5.8364, 3.8566, 2.0067, 0.5419, 0.5222 , 0.4994, 0.4572$, and  $0.4425$, respectively. After sampling $n=10,000$ values of $\theta_{q}$ for each of $q=1,2,\ldots, 9$, we used the resulting MAP estimates to compute the maximum log-likelihood for each $q$, the BIC evidence (\cite{kazianka2016}, \cite{minka2001}) given by 
\beq
\elabel{bic}
\pi(\vec{y}|q) \approx \left( \prod_{i=1}^{q} \widehat{\lambda}_{i} \right)^{-n/2}\left( \widehat{\sigma}^{2} \right)^{-n(d-q)/2} \, n^{-(k+q/2)}
\eeq
where $k=dq-q(q+1/2)$, and the Laplace evidence (\cite{kazianka2016}, \cite{minka2001}) given by
\beq
\elabel{laplace}
\pi(\vec{y}|q) \approx 2^{(k-q)/2} (\widehat{\sigma}^{2})^{n(d-q)} \, n^{-q/2} |A|^{-1/2} \prod_{i=1}^{q} \Gamma \left( \frac{d-i}{2}\right)
\eeq
for
$$
|A| = n^{k} \prod_{i=1}^{q} \prod_{j=i+1}^{d} (\widetilde{\lambda}_{j}^{-1}- \widetilde{\lambda}_{j}^{-1}) (\widehat{\lambda}_{i}-\widehat{\lambda}_{j})
$$
with
$$
\widetilde{\lambda}_{i} = \left\{
\begin{array}{lcl}
\widehat{\lambda}_{i} &,& \mbox{if} \,\, i \leq q\\
\frac{1}{d-q} \sum_{j=q+1}^{d} \widehat{\lambda}_{i} &,& \mbox{if} \,\, q < i \leq d.
\end{array}
\right.
$$
The idea would be to select models that maximize (\ref{e:bic}) and/or (\ref{e:laplace}). We refer the interested reader to \cite{kazianka2016} and \cite{minka2001} for more information about these criteria.

\begin{figure}[!htb]
\minipage{0.32\textwidth}
  \includegraphics[width=\linewidth]{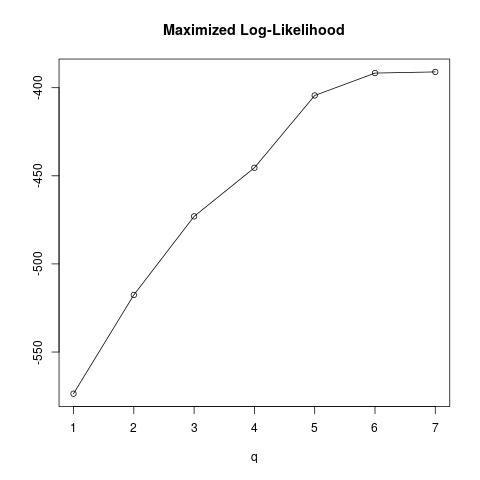}
\endminipage\hfill
\minipage{0.32\textwidth}
  \includegraphics[width=\linewidth]{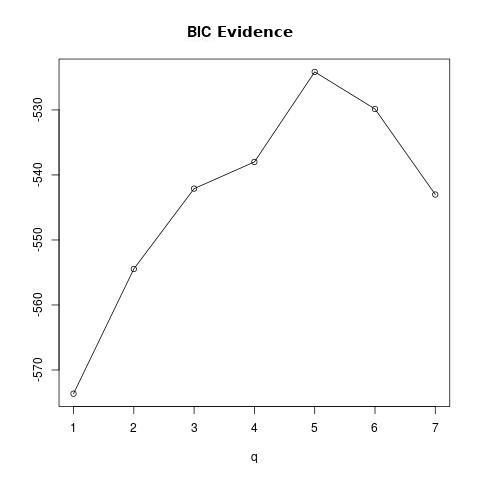}
\endminipage\hfill
\minipage{0.32\textwidth}%
  \includegraphics[width=\linewidth]{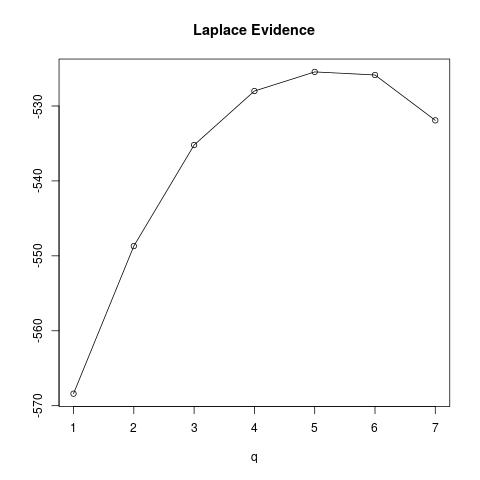}
\endminipage
\caption{Scoring the models}
\label{fig:score}
\end{figure}

The results are given in Figure \ref{fig:score} where, as expected, the maximized log-likelihood is increasing in $q$. The BIC appears to choose $q=5$ principal components, which makes sense given the variances ($10,8,6,4,2,0.5,0.5,0.5$) used to simulate the data. The Laplace evidence also chooses $q=5$, though it is not as pronounced as the BIC result. The average BCT in $10,000$ draws for this 8-dimensional model, with $\alpha=2$ and $\beta=3$, was 38 time steps. 

\section{Conclusion} 

The perfect OCNID Gibbs algorithm appears to be performing quite well for all of the examples we have tried, including the problem of order selection in Bayesian principal components analysis. Although it makes intuitive sense that $X_{i,L}^{(-n)}$ getting close to $X_{i,U}^{(-n)}$  will tend to make $X_{j,L}^{(-n)}$ and $X_{j,L}^{(-n)}$ close for $j=i-1,i+1$, the upper and lower paths for each component may also move apart again in the next time step since, for example, $X_{j-1,L}^{(-n)}$ depends also on $X_{j-2,L}^{(-n)}$. However, if the upper and lower processes for all $m$ components become simultaneously close, it is clear that they will stay that way since they are driven by uniform draws over smaller and smaller intervals.

It remains to be proven if, and under what circumstances, the  upper and lower processes for all $m$ components become simultaneously close. We have amassed empirical evidence, beyond the examples given here, that this will be the case and have observed that it happens faster than what would be expected based on chance alone. 

We see changes in average backward coupling times consistent with different choices of parameters and distributions with heavier or lighter tails. For example,  in \Section{expsec}, where the distributions are all exponential, if the rate parameters are decreasing so that the minimum OCNID order statistic has the highest rate and the maximum has the lowest rate, it is not surprising that coupling is quite fast. However, for different orderings of parameters, coupling is slower, as expected, but not significantly so. It is our hope for the future to be able to prove that coupling will happen in finite time and to be able to say something about the coupling rate, as least for some classes of distributions, if not in general.

%%%%%%%%%%%%%%%%%%%%%%%%%%%%%%%%%%%%%%%%%%%%%%%%%%%%%%%%%%%%%%%%%%%%
%%%%%%%%%%%%%%%%%%%%%%%%%%%%%%%%%%%%%%%%%%%%%%%%%%%%%%%%%%%%%%%%%%%%

%\bibliographystyle{/home/student/schneidu/TEX/elsart-harv}
%\bibliographystyle{elsart-harv}
\bibliographystyle{plain}

\bibliography{masterbib}
%\bibliography{/home/corcoran/BIBS/masterbib}

\end{document}